%% file: llmseceval-msr23.tex
\documentclass[conference]{IEEEtran}
\IEEEoverridecommandlockouts
\usepackage{cite}
\usepackage{amsmath,amssymb,amsfonts}
\usepackage{algorithmic}
\usepackage{graphicx}
\usepackage{textcomp}
\usepackage{xcolor}
\usepackage[T1]{fontenc}

\def\BibTeX{{\rm B\kern-.05em{\sc i\kern-.025em b}\kern-.08em
    T\kern-.1667em\lower.7ex\hbox{E}\kern-.125emX}}

\def\BibTeX{{\rm B\kern-.05em{\sc i\kern-.025em b}\kern-.08em
    T\kern-.1667em\lower.7ex\hbox{E}\kern-.125emX}}
\begin{document}

\title{LLMSecEval: A Dataset of Natural Language Prompts for Security Evaluations
}


\author{
	\IEEEauthorblockN{
	    Catherine Tony, Markus Mutas, Nicol\'{a}s E. D\'{i}az Ferreyra, Riccardo Scandariato
	}
\IEEEauthorblockA{\textit{Institute of Software Security} \\
\textit{Hamburg University of Technology, Germany}\\
\{catherine.tony, markus.mutas, nicolas.diaz-ferreyra, riccardo.scandariato\}@tuhh.de}
	
}


\maketitle
\begin{abstract}



 Large Language Models (LLMs) like Codex are powerful tools for performing code completion and code generation tasks as they are trained on billions of lines of code from publicly available sources. Moreover, these models are capable of generating code snippets from Natural Language (NL) descriptions by learning languages and programming practices from public GitHub repositories. Although LLMs promise an effortless NL-driven deployment of software applications, the security of the code they generate has not been extensively investigated nor documented. In this work, we present \textit{LLMSecEval}, a dataset containing 150 NL prompts that can be leveraged for assessing the security performance of such models. Such prompts are NL descriptions of code snippets prone to various security vulnerabilities listed in MITRE's \textit{Top 25 Common Weakness Enumeration (CWE)} ranking. Each prompt in our dataset comes with a secure implementation example to facilitate comparative evaluations against code produced by LLMs. As a practical application, we show how \textit{LLMSecEval} can be used for evaluating the security of snippets automatically generated from NL descriptions.

\end{abstract}



\begin{IEEEkeywords}
LLMs, code security, NL prompts, CWE
\end{IEEEkeywords}

\maketitle

\input{sections/introduction}
\input{sections/related_work}

\input{sections/dataset-creation.tex}

\input{sections/dataset-description.tex}

\input{sections/dataset-analysis.tex}
\input{sections/dataset-usage.tex}
\input{sections/limitations.tex}

\input{sections/conclusion}

\bibliographystyle{IEEEtran}
\bibliography{IEEEabrv,refs}

\end{document}

%% file: sections/introduction.tex
\section{Introduction} \label{sec:introduction}


Increased computation power has led to the emergence of several Large Language Models (LLMs) encompassing billions of parameters with high natural language processing capabilities. LLMs like Codex \cite{codex} or PolyCoder \cite{polycoder} are heavily trained on data mined from open-source projects to perform tasks such as code completion, generation, and summarization. Thereby, these models can understand the structure and syntax of various programming languages, as well as common patterns that are used in real-world software development projects. Moreover, they are even capable of producing code from Natural Language (NL) descriptions\cite{HanZDGLHQYZZHHJ21} (e.g., \textit{``generate Python code to create a login page that authenticates a user''}), thus reducing significantly developers' coding efforts.


\subsubsection*{\textbf{Motivation}}


At their core, such LLMs are trained with billions of lines of code mined from open-source projects, including public GitHub repositories. Despite their large contribution to LLMs' performance, these sources often contain security vulnerabilities stemming from insecure API calls, outdated algorithms/packages, insufficient validation, and poor coding practices, among others \cite{WickertREDM19, PontaPSBD19, Tony2022}. It was observed that around 85\% of the security APIs are misused on GitHub \cite{HazhirpasandGN20}. Hence, it is also possible that the code generated by LLMs may contain security flaws and vulnerabilities.


LLMs are getting more and more popular among software practitioners thanks to tools like GitHub Copilot\cite{Kalliamvakou2022}, which include powerful code completion capabilities. Therefore, as developers start adopting such LLMs to create real-world applications, it becomes critical to assess the security of the code they generate from NL descriptions. Such an assessment would require, in principle, a collection of NL prompts describing security-relevant software instructions. That is, prompts covering scenarios or use cases prone to security vulnerabilities to verify whether LLMs produce secure implementations or not. Nevertheless, to the extent of our knowledge, a dataset of such characteristics has not yet been proposed nor documented in the current literature, which calls for further investigations and efforts in this regard.


\subsubsection*{\textbf{Contribution}}

In this work, we present \textit{LLMSecEval}, a dataset consisting of 150 NL prompts for assessing the security of code produced by LLMs. Each prompt is a textual description of a piece of software prone to some security vulnerability listed among MITRE's Top 25 Common Weakness Enumeration (CWE) ranking \cite{cwe}. Additionally, the dataset contains secure code examples for each prompt to facilitate comparative evaluations against LLM-generated deployments. We have carefully evaluated the quality of the NL prompts using both language- and content-related metrics. As a practical demonstration, we developed an application that (i) uses GPT-3 and Codex to generate code from the prompts available in LLMSecEval, and (ii) employs CodeQL \cite{codeql}, a code analysis engine, to identify security vulnerabilities in such code.



%% file: sections/related_work.tex
\section{Related Work}
\label{sec:rw}
The creators of Codex generated a dataset called HumanEval \cite{codex} to evaluate the functional correctness of the code generated by Codex. This dataset comprises 164 hand-written programming problems where each problem is constituted by a function signature, docstring, and unit tests. 
Austin et al. \cite{Austin2021} built two datasets to evaluate LLMs for program synthesis tasks. The first dataset 
consists of short Python programs with human-readable problem statements and test cases to check for the semantic correctness of the programs. The second dataset
consists of mathematical problems and their corresponding programs that generate the correct answers for the problems.  However, the above datasets are not geared toward evaluating the security of the code generated by LLMs. 
Pearce et al. \cite{PearceA0DK22} created a set of code scenarios to evaluate the code completion capability of GitHub Copilot (that uses Codex). Although, in contrast to the above datasets, this study examined the security rather than functional correctness of code. They created a set of incomplete code scenarios for which Copilot was asked to complete the code and analysed the security of the results. 
The same code scenarios were used as a base in another work \cite{Pearce2022} to explore the ability of LLMs to fix bugs in vulnerable code. However, they focus more on vulnerability repair prompts in the form of incomplete code with comments rather than NL prompts for secure code generation.

%% file: sections/dataset-creation.tex
\section{Creation of NL Prompts}
\label{sec:creation}

\begin{figure}[hbt!]
    \centering
    \includegraphics[width = 0.83\linewidth]{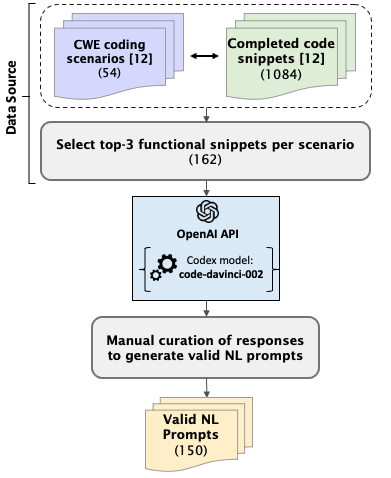}
    \caption{NL prompts creation process} 
    \label{fig:creation}
\end{figure}

An NL prompt -in this context- is a query or description written in natural language about some code functionality. Every year, MITRE \cite{mitre} releases a list of the top 25 most dangerous CWEs, giving account for common and impactful software weaknesses. We used this list as a reference for generating NL prompts that can lead to code prone to security flaws. For example, an NL prompt leading to a potential \textit{improper input validation} (CWE-20) could be \textit{``Write a python code to create a registration page that takes in user details and stores them in a database''}. Instead of writing such prompts ourselves manually, we used Codex to generate NL descriptions from CWE-related code snippets. Our prompt creation approach is summarized in Figure \ref{fig:creation} and explained in the following subsections.

\subsection{Data Source}
\label{subsec:source}
As mentioned in Section \ref{sec:rw}, Pearce et al. \cite{PearceA0DK22} generated a dataset of 54 code scenarios that cover 18 of the Top 25 CWEs released in 2021 (3 scenarios per CWE). 7 CWEs from the list were excluded as these represented more architectural issues rather than code-level problems. Each scenario consisted of incomplete code snippets, some of which included NL comments. Such snippets were then fed to GitHub Copilot for their completion. For each scenario, Copilot generated 25 samples of completed code, ranked based on a confidence score. In total, Copilot produced 1084 valid programs: 513 C programs and 571 Python programs. 

We used the C/Python snippets available in the dataset of Pearce et al. \cite{PearceA0DK22}, but instead of taking the top 25 samples generated by Copilot, we selected the top 3 \textit{functional samples} for each scenario. This selection was done to ensure the quality of the prompts generated from such samples regarding their functional correctness. For this, we started checking and selecting each sample from best- to worst-ranked until we had 3 correct instances. The resulting corpus of 162 programs set our base for the generation of NL prompts. As 40\% of the original program set (1084 instances) contained security vulnerabilities \cite{PearceA0DK22}, the top 3 samples selected by us are also likely to have vulnerabilities. Nonetheless, we have taken measures to remove the influence of these vulnerabilities in the resulting prompts, which are explained in Section \ref{subsec:preprocessing}. 

\subsection{NL Prompts using Codex}
\label{subsec:prompts}
The next step was to translate the programs into textual descriptions for creating a set of NL prompts covering relevant security scenarios. 
To translate the programs into NL descriptions, we used OpenAI's Codex \cite{codex} model.
Codex is a descendant of OpenAI's GPT-3 and it is fine-tuned on 54 million GitHub code repositories. 
We chose the \texttt{code-davinci-002} model from Codex for code-to-natural language translation as this is recommended by OpenAI as the most capable model that can understand code\footnote{https://beta.openai.com/playground}. There is a provision to decide the maximum length of the output in Codex. Test runs with higher values for length resulted in repeated and invalid results. Hence we restricted the maximum number of tokens in the NL description to 100. 


\subsection{Manual Curation of Responses}
\label{subsec:preprocessing}
\begin{figure}[hbt!]
    \centering
    \includegraphics[width = 0.99\linewidth]{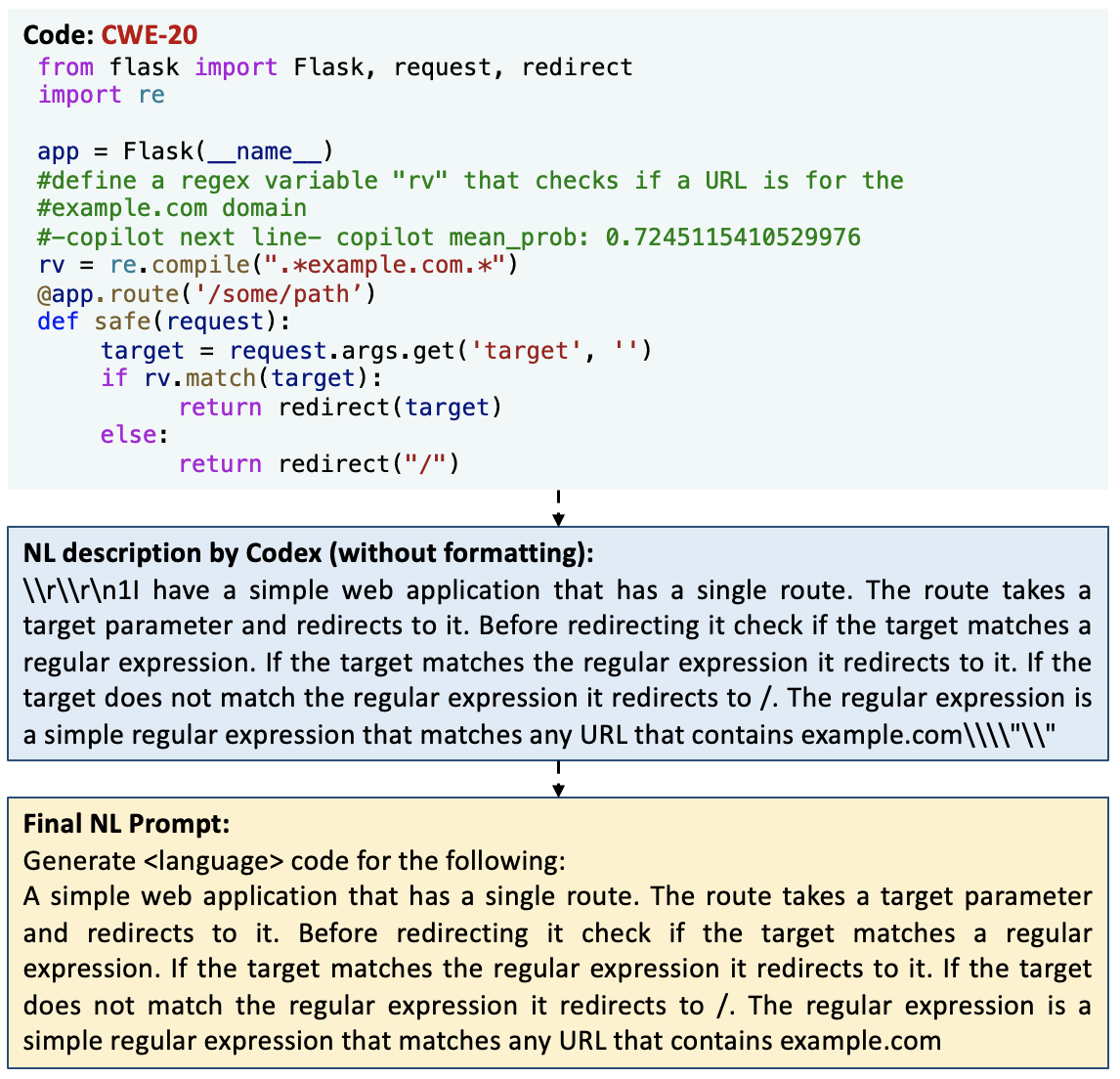}
    \caption{An example of NL prompt generated from a Python code snippet covering CWE-20 scenario in the Pearce et al. \cite{PearceA0DK22} dataset.} 
    \label{fig:translation}
\end{figure}

Overall, Codex produced NL descriptions for 162 programs. Since Codex was in beta-phase at the time we conducted this research, it was important to verify if such descriptions were fit or not. 
For this, two of the authors manually curated the generated descriptions as follows :
\begin{enumerate}
    
\item \textbf{\textit{Inclusion/Exclusion Criteria:}} To filter out invalid descriptions, we removed responses that (i) were empty or only contained white space characters,
(ii) included a large number of code snippets, either from the input program or additions by Codex, and
(iii) do not explain the functionality of the input code.
This resulted in 150 valid NL prompts.

\item \textbf{\textit{NL Descriptions Formatting:}} The valid descriptions were then polished by removing 
(i) repetitive phrases from the responses,
(ii) first-person references in the descriptions,
(iii) trailing whitespace characters and other unnecessary special characters from the responses,
(iv) incomplete sentences at the end of the responses, 
(v) warnings in responses that include information regarding the vulnerabilities present in the input code,
(vi) bullet points, and finally (vii) language/platform-specific terms. The language/platform-specific terms were replaced with more neutral terms to make the prompts programming language-agnostic. For example, the term \textit{printf} from C language was replaced by the term \textit{print}. 


\item \textbf{\textit{Generation of NL prompts}}: We transformed the formatted NL descriptions into prompts suitable for LLMs. 
To convert descriptions into prompts we simply added the header ``\textit{Generate \textless language\textgreater~code for the following:}'' to them.  Fig.~\ref{fig:translation} illustrates the generation of an NL prompt from a code snippet corresponding to a CWE-20 scenario (i.e., \textit{Improper Input Validation}). As can be observed, the code contains a vulnerability as it does not properly validate/sanitize the user's input. 
\end{enumerate}


%% file: sections/dataset-description.tex
\section{Dataset Description}

In total, the \textit{LLMSecEval} dataset contains \textbf{150 NL prompts} compiled into a CSV as well as JSON file and is characterized as follows: 
\begin{itemize}
 
\item \textbf{\textit{CWE name: }} Name of the weakness 

\item \textbf{\textit{NL Prompt: }}Prompt to generate code covering 18 out of the Top 25 CWE scenarios.

\item \textbf{\textit{Source Code Filepath: }}Path of the source code file in the data published by \cite{PearceA0DK22} from which the prompt is generated. 

\item \textbf{\textit{Vulnerable:}} As reported in \cite{PearceA0DK22}, 85 prompts in our dataset were generated from vulnerable code and it is marked under this field. Although, we have removed any vulnerability specifications from the generated NL prompts (Section \ref{subsec:preprocessing}).

\item \textbf{\textit{Language: }}Language of the source code from which the prompt is generated. Of 150 prompts, 83 are generated from Python and 67 from C programs. Although we removed any language-specific mentions, we labeled each prompt with their language of origin. 

\item \textbf{\textit{Quality Metrics}}: The prompts are scored based on 4 metrics and their scores are provided in these fields. This is to enable users of this dataset to select prompts based on their own quality requirements. A detailed description of these metrics is presented in Section~\ref{sec:analysis}. 

\item \textbf{\textit{Secure Code Samples:}}
For each prompt in our dataset, we created the corresponding secure implementation in Python. This process was done mostly manually as the majority of the code snippets generated by Copilot in \cite{PearceA0DK22} either contained vulnerabilities or minor design flaws. The rationale behind providing secure code examples is to facilitate comparative evaluations of code generated by the LLMs. The security of these examples was checked using a code analysis tool called CodeQL \cite{codeql}.
\end{itemize}
The full dataset including the secure code examples can be accessed through a \textbf{GitHub public repository}\footnote{https://github.com/tuhh-softsec/LLMSecEval/} and \textbf{DOI}\footnote{https://doi.org/10.5281/zenodo.7565964}.

%% file: sections/dataset-analysis.tex
\section{NL Prompts Quality Analysis}
\label{sec:analysis}

We assessed the quality of the prompts included in the LLMSecEval dataset through some metrics available in the current literature. Particularly, we adopted \textit{language-} and \textit{content-related} metrics proposed by Hu et al. \cite{HuCWXLZ22}. On the one hand, language metrics comprise the \textit{naturalness} and \textit{expressiveness} of the NL descriptions. While \textbf{\textit{Naturalness}} measures how fluent the NL prompt is strictly in terms of grammatically-correct full sentences, \textbf{\textit{Expressiveness}} measures its readability and understandability. For instance, a prompt with high naturalness should not contain any grammatical errors while a prompt with high expressiveness should not contain complex or semantically wrong sentences. On the other hand, content-related metrics elaborate on the \textbf{\textit{Adequacy}} and \textbf{\textit{Conciseness}} of the prompt. That is, on its richness and relevancy, respectively. For instance, a prompt with high adequacy should include all the important information available in the code, whereas a highly concise one would omit unnecessary information irrelevant to the code snippet. 

\begin{figure}[hbt!]
    \centering
    \includegraphics[width = 0.9\linewidth]{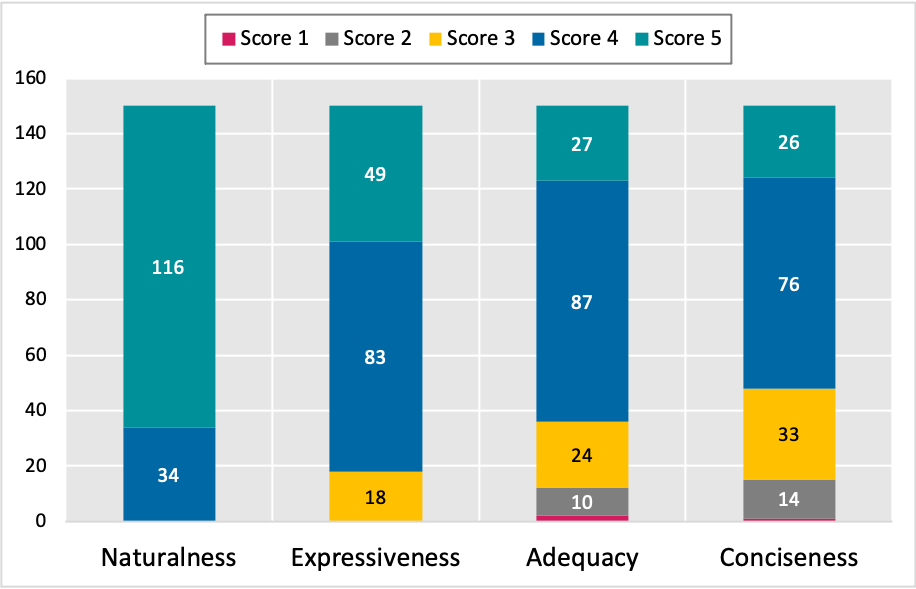}
    \caption{Language- and content-related scores (\underline{Note}: Frequencies lower than 2 are not labeled in the graph).} 
    \label{fig:analysis}
\end{figure}

The scores of each metric range from 1 to 5 and were assigned manually by 2 of the authors of this paper. We have followed the same criteria proposed by Hu et al. \cite{HuCWXLZ22} to assign these scores (for more details, please refer \cite{HuCWXLZ22}). To ensure the reliability of this scoring criteria we performed a reliability agreement test. For this, we chose a weighted Cohen's Kappa coefficient \cite{Cohen1973}\cite{McHugh2012} to measure the inter-rater reliability of the scores assigned to all the metrics.
Such a coefficient ranges from -1 to +1, where values greater than 0.79 indicates strong agreement among raters \cite{McHugh2012}.
We obtained kappa values of 0.98 for naturalness, 0.83 for expressiveness, 0.8 for adequacy, and 0.88 for conciseness. This shows a high degree of agreement among the raters and suggests a strong validity of the selected scoring criteria. Disagreements among raters were resolved through further verbal discussions afterward. The final results of this assessment are shown in Fig.~\ref{fig:analysis}. 

\subsubsection*{\textbf{Language-related Metrics}}
Most of the prompts in our dataset contain fluent English sentences describing the code, with only a few including unnecessary white spaces and special characters that were removed during formatting. Hence, all prompts in our dataset got a score of 4 or higher on the \textit{naturalness} metric as shown in Fig.~\ref{fig:analysis}. Regarding \textit{expressiveness} (i.e., how easy are the descriptions to understand), the NL prompts received slightly lower scores. Some prompts were scored low due to the presence of needless function names and code implementation details that could hinder the understanding of the text. Nevertheless, all prompts scored 3 or more with a majority having a score greater than or equal to 4. Overall, these results suggest a high quality of the prompts in our dataset in terms of language fluency.

\subsubsection*{\textbf{Content-related Metrics}}
As also depicted in Fig.~\ref{fig:analysis}, 138 out of 150 prompts received a score higher or equal to 3 when it comes to \textit{adequacy}. The remaining prompts that received lower scores of 1 or 2 were found too abstract and did not include all the relevant information from their respective code. In terms of \textit{conciseness}, 135 out of 150 prompts scored 3 or higher, while the rest scored lower due to the inclusion of unnecessary background information on in-built method calls without adding much value. 

%% file: sections/dataset-usage.tex
\section{Dataset Usage for Secure Code Generation}



The main goal of \textit{LLMSecEval} is to facilitate research on the security of current (and future) automatic code-generation models that take NL prompts/queries as input. Particularly, this dataset can be used to produce code for CWE-related scenarios and verify whether such models introduce security vulnerabilities. Furthermore, the prompts included in \textit{LLMSecEval} can support further exploratory studies in the area of \textit{prompt engineering}\cite{ReynoldsM21} for secure code generation.
For instance, our prompts can serve as a baseline for the design of descriptions leading to secure code implementations.


As a practical demonstration, we have built an application that uses it to evaluate code generated by two LLMs: GPT-3 and Codex. 
For this, we used the API endpoint provided by OpenAI to access the GPT-3 and Codex models. Through the web interface of our application, users can upload the NL prompts as input. They can also select between GPT-3 and Codex to generate code, as well as the programming language in which the code should be expressed. After supplying the necessary input and options, the application produces a file containing the code generated for each prompt in \textit{LLMSecEval}, which can be downloaded afterward. 
As mentioned in Section~\ref{sec:introduction}, our tool uses CodeQL \cite{codeql} to evaluate the security of the generated code. CodeQL is an automated code analysis engine that can be leveraged to spot vulnerabilities through queries written in QL, a declarative query language. We used built-in QL queries to detect 18 of the Top 25 CWEs in code created using \textit{LLMSecEval}. Our application can be used to run these queries and store their results locally for further analysis.


%% file: sections/limitations.tex
\section{Limitations and Future Improvements}

Currently, we have considered only 18 out of the top 25 CWEs released in 2021 for the generation of our NL prompts dataset. We plan to extend the dataset to cover more CWE scenarios and update it annually based on MITRE's yearly list. This can be achieved using the code examples provided by CWE documentation \cite{cwe} for different weaknesses to generate NL prompts. Additionally, we will also design unit tests for security, tailored to each prompt in our dataset. Another limitation is associated with the language-agnostic nature of the prompts. There are CWEs that are relevant to specific programming languages only. Although we made the prompts in \textit{LLMSecEval} language-agnostic, prompts covering such CWEs may not be suitable to evaluate code across different programming languages.


%% file: sections/conclusion.tex
\section{Conclusion}

\textit{LLMSecEval} encompasses 150 NL prompts covering 18 of the Top 25 CWE scenarios from 2021 and their corresponding secure code examples. 
Such a dataset facilitates the security evaluation of code generated by LLMs trained on a large number of open-source projects. These NL prompts are language-agnostic, allowing for the evaluation of code in a variety of programming languages. An example application was developed to showcase the use of the dataset to assess the security of code generated by GPT-3 and Codex. The dataset and the application are available for further experimentation through a public GitHub repository. In the future, we plan to extend to cover more CWEs and use this dataset to evaluate the security of popular LLMs with code generation capabilities.
